\documentclass[lettersize,journal]{IEEEtran}
\usepackage{amsmath,amsfonts}
\usepackage{algorithmic}
\usepackage{algorithm}
\usepackage{array}
\usepackage[caption=false,font=normalsize,labelfont=sf,textfont=sf]{subfig}
\usepackage{textcomp}
\usepackage{stfloats}
\usepackage{url}
\usepackage{verbatim}
\usepackage{graphicx}
\usepackage{cite}
\usepackage{orcidlink} 
\usepackage{url}            
\usepackage{booktabs}       
\usepackage{amsfonts}       
\usepackage{nicefrac}       
\usepackage{microtype}      
\usepackage{xcolor}         
\usepackage{multirow}
\usepackage{makecell}
\usepackage{colortbl}
\usepackage{tcolorbox}
\usepackage{caption}

\usepackage{cite}
\usepackage{algorithm}
\usepackage{algorithmic}

\usepackage{enumitem}
\setlist{nolistsep}

\usepackage{amsmath, amsthm}
\usepackage{amssymb}
\usepackage{mathtools}
\newtheorem{theorem}{Theorem}
\newtheorem{lemma}{Lemma}
\newtheorem{definition}{Definition}
\newtheorem{assumption}{Assumption}

\newtheorem{remark}{Remark}
\newtheorem{problem}{Problem}

\newtheorem{corollary}{Corollary}

\definecolor{cvprblue}{rgb}{0.21,0.49,0.74}
\newcommand{\revision}[1]{\textcolor{black}{#1}}

\newcommand{\eqautoref}[1]{\hyperref[#1]{Equation (\ref{#1})}}
\newcommand{\ineqautoref}[1]{\hyperref[#1]{Inequality (\ref{#1})}}

\newcommand{\algautoref}[1]{\hyperref[#1]{Algorithm~\ref{#1}}}
\newcommand{\apdxautoref}[1]{\hyperref[#1]{Appendix~\ref{#1}}}

\newcommand{\formulaautoref}[1]{\hyperref[#1]{Formula~\ref{#1}}}
\newcommand{\problemautoref}[1]{\hyperref[#1]{Problem~\ref{#1}}}

\newcommand{\coroautoref}[1]{\hyperref[#1]{Corollary~\ref{#1}}}

\newcommand{\lineautoref}[1]{\hyperref[#1]{line~\ref{#1}}}
\begin{document}

\title{Optimal Client Sampling in Federated Learning with Client-level Heterogeneous Differential Privacy}

\author{
Jiahao Xu\orcidlink{0000-0002-4888-252X},~\IEEEmembership{Graduate Student Member,~IEEE}, Rui Hu\orcidlink{0000-0003-3317-1765},~\IEEEmembership{Member,~IEEE}, Olivera Kotevska\orcidlink{0000-0003-1677-2243},~\IEEEmembership{Senior Member,~IEEE}
\thanks{The work of Jiahao Xu and Rui Hu was supported by the National Science Foundation under the Harnessing the Data Revolution for Nevada Fire Science (HDRFS) Seed Grant NSHE-24-37. This material is based upon work co-supported by the U.S. Department of Energy, Office of Science, Office of Advanced Scientific Computing Research under Contract No. DE-AC05-00OR22725. This manuscript has been co-authored by UT-Battelle, LLC under Contract No. DE-AC05-00OR22725 with the U.S. Department of Energy. The United States Government retains and the publisher, by accepting the article for publication, acknowledges that the United States Government retains a non-exclusive, paid-up, irrevocable, world-wide license to publish or reproduce the published form of this manuscript, or allow others to do so, for United States Government purposes. The Department of Energy will provide public access to these results of federally sponsored research in accordance with the DOE Public Access Plan (http://energy.gov/downloads/doe-public-access-plan).}
\thanks{Jiahao Xu and Rui Hu are with the Department of Computer Science and Engineering, University of Nevada, Reno, Reno, NV 89557, USA (e-mail: jiahaox@unr.edu; ruihu@unr.edu). Olivera Kotevska is with the Oak Ridge National Laboratory, Oak Ridge, TN 37830, USA (e-mail: kotevskao@ornl.gov). This work was partially done during Jiahao's internship at the Oak Ridge National Laboratory. Rui Hu is the corresponding author.}%
\thanks{The code is available at \url{https://github.com/JiiahaoXU/GDPFed}.}
}

\markboth{Journal of \LaTeX\ Class Files,~Vol.~14, No.~8, August~2021}%
{Shell \MakeLowercase{\textit{et al.}}: A Sample Article Using IEEEtran.cls for IEEE Journals}


\maketitle

\setcounter{page}{1}
\setcounter{section}{0}

\begin{abstract}
Federated Learning with client-level differential privacy (DP) provides a promising framework for collaboratively training models while rigorously protecting clients’ privacy. However, classic approaches like DP-FedAvg struggle when clients have heterogeneous privacy requirements, as they must uniformly enforce the strictest privacy level across all clients, leading to excessive DP noise and significant degradation in model utility. Existing methods to improve the model utility in such heterogeneous privacy settings often assume a trusted server and are largely heuristic, resulting in suboptimal performance and lacking strong theoretical foundations. In this work, we address these challenges under a practical attack model where both clients and the server are honest-but-curious. We propose GDPFed, which partitions clients into groups based on their privacy budgets and achieves client-level DP within each group to reduce the privacy budget waste and hence improve the model utility. Based on the privacy and convergence analysis of GDPFed, we find that the magnitude of DP noise depends on both model dimensionality and the per-group client sampling ratios. To further improve the performance of GDPFed, we introduce GDPFed$^+$, which integrates model sparsification to eliminate unnecessary noise and optimizes per-group client sampling ratios to minimize convergence error. Extensive empirical evaluations on multiple benchmark datasets demonstrate the effectiveness of GDPFed$^+$, showing substantial performance gains compared with state-of-the-art methods. 
\end{abstract}


\section{Introduction}\label{sec: intro}

\IEEEPARstart{T}{raditional} \revision{centralized Machine Learning (ML) frameworks require collecting all training data at a single node (e.g., a central server), which raises serious privacy concerns and communication burdens. These issues are particularly evident in Internet of Things (IoT) systems, where massive numbers of resource-constrained devices continuously generate sensitive data and transmit it over bandwidth-limited networks. To address these challenges, Federated Learning (FL)~\cite{fedavg} has emerged as a distributed ML paradigm that enables collaborative model training across decentralized data sources without centralizing raw data. FL has been widely applied in IoT-related domains such as healthcare~\cite{healthcare} and remote sensing~\cite{remotesense}, where data privacy, communication efficiency, and on-device computation are critical. In a typical FL system, multiple local clients, such as IoT or edge devices, collaboratively train a shared global model under the coordination of a central server. In each training round, the server broadcasts the global model to a subset of clients, which update the model using their local data. The resulting model updates are then sent back to the server and aggregated to refine the global model. This iterative process continues until global model convergence.}

Although the FL paradigm keeps sensitive training data on clients, recent studies have shown that adversaries can still infer private information through well-crafted inference attacks~\cite{shokri2017membership, nasr2019comprehensive, mem1, mem2, mem3}. To mitigate privacy risks, \textit{differential privacy} (DP)~\cite{dp_og}, a widely adopted standard for incorporating formal privacy guarantees, has been integrated into the FL algorithm~\cite{dp-fedavg}. In the context of FL, DP can be applied at two distinct protection levels: \textit{record-level DP}, which protects individual data points within a client's dataset, and \textit{client-level DP}, which protects the participation of a client (i.e., the client's entire dataset). This work focuses on achieving \textit{client-level differentially private FL} (DPFL), as it typically yields better model utility than its record-level counterpart in \textit{cross-device} settings~\cite{fedsmp}. 
In the literature, client-level DPFL is typically implemented using the Gaussian mechanism~\cite{dp_og}, where each client's model update is perturbed by adding Gaussian noise scaled according to a \textit{uniform} privacy budget $\epsilon$ across all clients~\cite{andrew2021differentially, boenisch2023have, dp-fedsam}. A smaller $\epsilon$ provides stronger privacy guarantees but requires injecting larger noise, which consequently leads to more severe model utility degradation. 
These perturbed model updates are typically aggregated using \textit{secure aggregation} (e.g.,~\cite{bonawitz2017practical}), which cryptographically ensures that the server can only access their sum without observing individual contributions. This dual protection yields a differentially private aggregated model update that prevents client-level privacy inference even with an adversarial server.

However, in practice, clients often have heterogeneous privacy preferences, necessitating support for \textit{heterogeneous DP} (HDP)~\cite{jorgensen2015conservative}. In the literature, \cite{PFA} formally introduced the problem of \textit{DPFL with heterogeneous privacy requirements} for different clients (HDPFL), where each client naturally has an individual privacy budget reflecting their privacy needs. In this setting, ensuring record-level HDP is relatively straightforward, and numerous studies have proposed to improve model utility~\cite{boenisch2023have, shen2023pldp, yang2021federated, ma2025power, pmlr-v235-malekmohammadi24a, ccs_rdp}. In contrast, client-level HDPFL remains underexplored. To achieve client-level DP with heterogeneous privacy requirements, conventional approaches such as DP-FedAvg~\cite{dp-fedavg} \textit{must} satisfy the most stringent privacy budget among all clients, which severely limits overall model utility. A more practical alternative partitions clients into groups and enforces client-level DP at the group level. To improve the model utility in this scenario, recent efforts include manually adjusting per-group client sampling ratios~\cite{kiani2025differentially}, adjusting per-group training rounds~\cite{chathoth2021federated}, and mitigating the influence of noisy per-group updates~\cite{PFA}. However, these approaches assume a fully trusted server, which is often unrealistic in settings that are vulnerable to privacy inference attacks. Moreover, they primarily rely on heuristic methods without rigorous theoretical analysis to optimize the privacy-utility trade-off.

In this work, we aim to optimize the model utility in client-level HDPFL under a strong attack model where both the clients and the server are adversaries.
\revision{We propose \textbf{\emph{\underline{G}}}roup-based \textbf{\emph{\underline{D}}}ifferentially \textbf{\emph{\underline{P}}}rivate \textbf{\emph{\underline{Fed}}}erated Learning (GDPFed), a novel client-level HDPFL approach. In GDPFed, clients are grouped according to their privacy budgets, and client-level DP is enforced at the group level by using the minimum privacy budget within each group rather than the global minimum across all clients.}
This design preserves high model utility while accommodating heterogeneous privacy preferences across clients.
Building on this, we theoretically investigate how to maximize model utility in GDPFed while maintaining rigorous privacy guarantees. Through privacy and convergence analyses, we identify two key factors influencing convergence errors under fixed privacy budgets: (1) the model dimensionality, since DP noise is injected into each model parameter, causing the total noise to scale with model size; and (2) the per-group client sampling ratio, which provides privacy amplification effects on the overall guarantees. To reduce dimensionality-induced noise, we incorporate model sparsification into GDPFed, which eliminates less significant model parameters for each group with minimal utility drop. We then optimize the per-group client sampling ratios towards minimizing the convergence error, which extends GDPFed to GDPFed$^+$ with improved model utility. In summary, we make the following contributions:

\begin{itemize}[itemsep=5pt]
    \item We propose GDPFed, a novel \textit{client-level} HDPFL algorithm for environments where both server and clients are \textit{honest-but-curious}. GDPFed is specifically designed to improve model utility when clients have heterogeneous privacy preferences. By achieving client-level DP at a group-wise level, our approach mitigates privacy budget waste inherent in HDP settings, improving the model utility. GDPFed builds upon FedAvg framework, enabling seamless integration into existing FL systems.
    \item To further improve the model utility, while preserving the privacy guarantees, we propose GDPFed$^+$, which integrates per-group model sparsification into GDPFed and optimizes the per-group client sampling ratios to minimize the impact of DP noise on the model utility. \textit{To the best of our knowledge, this is the first work that optimizes client sampling ratios to enhance the privacy-utility trade-off in client-level HDP settings.}
    \item We conduct extensive evaluations on multiple benchmark datasets of DPFL, thoroughly comparing our methods against state-of-the-art baseline methods. The results consistently demonstrate that GDPFed outperforms existing methods in HDP settings, while GDPFed$^+$ further improves the model utility under the same privacy guarantee.  
\end{itemize}

The remainder of this paper is organized as follows. 
\autoref{sec: preliminary} introduces the system settings and preliminaries on DP.
\autoref{sec: related_work} reviews related work.
\autoref{sec: gdpfed} presents the proposed GDPFed method and its privacy analysis.
\autoref{sec: gdpfedplus} describes the enhanced GDPFed$^+$ method and provides a rigorous convergence analysis.
\autoref{sec: evaluation} reports comprehensive experimental results.
Finally, \autoref{sec: conclusion} concludes the paper.

\section{System Settings and Preliminary} \label{sec: preliminary}

\subsection{Attack Model} To achieve client-level DP, the literature typically assumes that the adversary is either \textit{honest-but-curious} clients~\cite{boenisch2023have, kiani2025differentially, dp-fedavg, PFA} or, in a stronger setting, both the clients and the server~\cite{fedsmp, dp-fedsam, fereidooni2021safelearn, kadhe2020fastsecagg}. In this work, we consider the latter, more challenging one. Specifically, the adversary is assumed to follow the prescribed training protocol honestly but remains curious about the private data of a target client, attempting to infer it from the shared messages. In addition, certain clients may collude with the server or with one another to extract sensitive information about a specific victim client. Moreover, the adversary may also take the form of a passive external eavesdropper who can observe all shared messages during training but does not actively inject false messages or disrupt communication~\cite{fedsmp}.

\subsection{Federated Learning and FedAvg} In a typical FL system, a set of~$ n $ clients aim to collaboratively train a shared global model $ \theta \in \mathbb{R}^d $ in an iterative manner under the coordination of a central server. Generally, the FL problem can be formulated as $\min_{\theta} (1/n)\sum_{i=1}^n f_i(\theta)$, where $f_i(\theta)= \mathbb{E}_{(z, y)\in D_i}l(\theta; z, y)$ represents the local learning objective of client $i$. Here, $l(\cdot)$ is the loss function, and $(z, y)$ is a datapoint sampled from the local dataset $D_i$ of client $ i\in[n]$. The classic method to solve the FL problem is known as \textit{Federated Averaging} (FedAvg)~\cite{fedavg}. Specifically, in each training round $t$, the server randomly selects a set of $r$ clients $\mathcal{S}^t$ with a client sampling ratio $q \in (0, 1]$ without replacement to participate in the local training. Each client $i\in\mathcal{S}^t$ then downloads the latest global model $\theta^{t-1}$ from the server, refines the model for $\tau$ iterations towards optimizing its local objective to obtain an updated local model $\theta_i^t$ and then sends its local model updates $\Delta_i^t = \theta_i^t - \theta^{t-1}$ back to the server. The server refines the global model by averaging the local updates as $\theta^{t} = \theta^{t-1}+ (1/r)\sum_{i \in \mathcal{S}^t} \Delta_i^t$. This process repeats for enough $T$ rounds to ensure that the global model converges. Since the server receives individual model updates from clients in each round, it poses a significant privacy risk, as a curious server can infer sensitive information from these updates~\cite{nasr2019comprehensive}.

\subsection{Differential Privacy} \label{sec: def_of_DP} The DP mechanism~\cite{dp_og, mironov2017renyi}, especially the Gaussian mechanism (see the formal definition in Lemma 6 in Appendix~A\revision{)}, has been employed as a rigorous approach for mitigating privacy threats in FL~\cite{blur_lus, fedsmp, dp-fedsam}. 
We give the formal definition of classic \((\epsilon, \delta)\)-DP in \autoref{DP}.
\begin{definition}[$(\epsilon,\delta)$-DP~\cite{dp_og}]\label{DP} 
Given privacy budget $\epsilon >0$ and failure parameter $0\leq \delta < 1$, a randomized mechanism $\mathcal{M}$ satisfies $(\epsilon,\delta)$-DP if for any two adjacent datasets $D, D^{\prime}$, any subset of outputs ${O} \subseteq \text{range}(\mathcal{M})$ satisfies
$
\Pr[\mathcal{M}(D) \in {O}] \leq e^{\epsilon} \Pr[\mathcal{M}(D^{\prime}) \in {O}] + \delta.
$
\end{definition}
In this work, as we consider client-level DP, we define the \textit{adjacent datasets} by adding or removing the \textit{entire} local dataset of a client in FL. The privacy budget $\epsilon$ defines the upper bound on privacy loss in a DP mechanism. A smaller $\epsilon$ indicates stronger privacy protection but requires injecting more intense noise into the learning process, which can significantly impact performance. Additionally, the failure parameter $\delta$ quantifies the probability that the DP guarantee may be violated. 

\section{Related Work} \label{sec: related_work}

\subsection{Client-level DP-FedAvg} \label{sec: related_CLDPFEDAVG} Compared with record-level DP~\cite{ccs_rdp, record3, han2025ppfl, 10416913}, which aims to protect every individual record in a client's dataset, client-level DP hides a single client's overall contribution. To achieve client-level DP under our attack model, one can use the DP-FedAvg algorithm~\cite{dp-fedavg}, which is presented in Algorithm~3 in Appendix~D. Specifically, before transmitting the local model update \(\Delta^t_i\) to the server at round $t$, each selected client clips its model update with a clipping threshold \(C\), and adds small amount of DP noise drawn from \(\mathcal{N}\left(0, C^2 \sigma^2/r \cdot \mathbf{I}^d\right)\), where \(\sigma^2\) is the noise multiplier. Notably, the noise multiplier $\sigma^2$ must be carefully calibrated to ensure that DP-FedAvg satisfies $(\epsilon, \delta)$-DP after $T$ training rounds. Theoretical analyses have established the relationship $\sigma^2 = \Omega(q^2 / \epsilon)$~\cite{abadi2016deep, mironov2019r}, implying that satisfying a smaller $\epsilon$ necessitates injecting larger noise. \revision{Furthermore, DP-FedAvg benefits from \textit{privacy amplification} via client subsampling~\cite{beimel2014bounds}, where each client is independently selected with probability $q$ in every training round.
}

After perturbing their updates locally, clients encrypt these noisy updates using a secure aggregation protocol (e.g., \cite{bonawitz2017practical}) and send them to the server. Secure aggregation is a commonly used practice in client-level DPFL~\cite{fedsmp, fereidooni2021safelearn, kadhe2020fastsecagg, dp-fedsam}, ensuring that a curious server only observes the aggregated sum of clients' updates, without access to individual contributions. In this setting, the aggregated model update received by the server is already perturbed with Gaussian noise  $\mathcal{N}\left(0, C^2 \sigma^2 \cdot \mathbf{I}^d\right)$. 
Finally, the global model is refined with the perturbed aggregated updates.  If the server is assumed to be trusted \cite{chathoth2021federated, kiani2025differentially, PFA}, these model clipping and perturbation operations can be directly applied to the aggregated model update on the server side to prevent clients from inferring private information.

The noise applied to model updates inherently reduces the utility of the global model. To mitigate this issue, numerous methods have been proposed, including model update regularization~\cite{andrew2021differentially, blur_lus, dp-fedsam} to ensure more robust local updates, optimized client sampling~\cite{aocs, delta, ribero2020communication} to select more informative clients, and sparsification~\cite{fedsmp, blur_lus, FedMPS, zhang2023byzantine} to remove unnecessary noise. \revision{For example, Wang et al.~\cite{FedMPS} sparsify each layer in model updates to remove unnecessary noise. However, these methods consider a homogeneous DP setting, where all clients share the same privacy preference.} In contrast, an HDP setting where clients have heterogeneous privacy preferences is more realistic and better aligned with practical deployment scenarios.

\subsection{FL with Heterogeneous Privacy Preferences} \label{sec: related_HDP} In practice, clients often have diverse privacy requirements due to varying policies or individual preferences, making it essential to consider FL under HDP~\cite{jorgensen2015conservative}. Liu et al.~\cite{PFA} first formalized the problem of HDPFL, allowing each client to specify a unique privacy budget that reflects their preferences. In this setting, record-level HDP is straightforward to implement by calibrating the DP noise individually per client~\cite{boenisch2023have, ccs_rdp, shen2023pldp, yang2021federated, PFA, ma2025power}. For example, Boenisch et al.~\cite{boenisch2023have} proposed IDP-FedAvg, which assigns data sampling ratios and clipping thresholds based on each record's privacy budget. \revision{Ma et al.~\cite{ma2025power} studied the client selection problem in FL with record-level HDP. However, achieving client-level HDP, where the goal is to protect a single client's contribution from being inferred, poses greater challenges.}

Standard approaches such as DP-FedAvg~\cite{dp-fedavg} in this heterogeneous setting have to calibrate noise to satisfy the most stringent privacy requirement among clients, leading to excessive noise for clients with more relaxed privacy preferences and thus poor model utility~\cite{kiani2025differentially}. A more privacy-efficient approach is to partition clients into groups based on their privacy budgets and ensure client-level DP within each group~\cite{kiani2025differentially, PFA, chathoth2021federated}. \revision{For instance,  Kiani et al.~\cite{kiani2025differentially} proposed a dynamic HDPFL framework where clients in different groups consume less privacy budget in early training rounds. 
While they formulate the client sampling ratio optimization problem, they manually tune each group's sampling ratio, limiting the method's theoretical rigor. Instead, our work is the first to provide a solution to the problem with theoretical justification.}
Another related method, Projected Federated Averaging (PFA)~\cite{PFA}, retains updates from groups with high privacy budgets while projecting updates from low-budget groups onto the principal subspace learned from the high-budget group. Compared to PFA, our method improves the privacy-utility trade-off through both theoretical analysis and optimization techniques.

\section{Federated Learning with Heterogeneous Group Client-Level DP}
\label{sec: gdpfed}

\subsection{GDPFed} 
\label{subsec:gdp_fedavg}
Recall that in this work, we consider an HDPFL setting where each client has its own privacy budget $\epsilon_i, \ \forall i\in[n]$. The objective is to collaboratively train a global model with satisfactory utility while respecting each client’s privacy preference. To achieve this, our proposed method, GDPFed, partitions all clients into $M$ groups $\mathcal{G}_1, \mathcal{G}_2, \dots, \mathcal{G}_M $ based on their privacy budgets. Note that the FL problem now is formalized as $\min_{\theta} \sum_{m \in [M]} \omega_m  \sum_{i\in \mathcal{G}_m}f_{m,i}(\theta)$,
where $f_{m,i}(\theta)$ is the local learning objective of client $i$ in $\mathcal{G}_m$ and $\omega_m$ is a reweighting parameter for each group. In each training round $t$ of GDPFed, the server samples a subset of $r_m$ clients $\mathcal{S}^t_m$ from each group $m \in [M]$ where the number of sampled clients $r_m$ in group $m$ is determined by the client sampling ratio $q_m$ and calculated as $r_m = q_m  |\mathcal{G}_m| $. To achieve client-level DP within each group, every local model update in group $m$ is perturbed by adding Gaussian noise drawn from $\mathcal{N}(0, C^2 \sigma_m^2 / r_m \cdot \mathbf{I}^d)$ after clipping with clipping threshold $C$. Note that the noise multiplier $\sigma_m^2$ is set to satisfy the minimum privacy budget within each group, denoted by $\epsilon_m = \min\{\epsilon_{m,i}\}_{i\in \mathcal{G}_m}$, to ensure that clients' privacy losses are smaller than their budgets. Consequently, selected clients send the perturbed local updates via secure aggregation. 
One can follow the approach in~\cite{bonawitz2019federated, fedsmp} to implement secure aggregation, and we note that designing a novel secure aggregation protocol is beyond the scope of this paper.
The server receives the model update summation from each group and aggregates them with reweighting parameters to refine the global model. This process will repeat for $T$ rounds to ensure that the global model achieves sufficient utility.


\subsection{Privacy Analysis of GDPFed} \label{sec: privacy_analysis}
Now we provide the detailed privacy analysis for the proposed GDPFed. Specifically, we first examine the conditions under which each group in GDPFed satisfies the DP requirements. Then, we present privacy guarantees for the entire GDPFed.
We provide per-group privacy guarantees of GDPFed in \autoref{thm:privacy_loss_per_group}.

\begin{theorem}[Per-Group Privacy Guarantees of GDPFed]
\label{thm:privacy_loss_per_group}
Suppose clients in group $m$ are sampled without replacement with probability $q_m$ at each round. For any $\epsilon_m < 2\log(1/\delta)$ and $\delta \in (0,1)$, GDPFed satisfies $(\epsilon_m, \delta)$-DP for clients in group $m$ after $T$ rounds if \[\sigma^2_m \geq 7q^2_mT(\epsilon_m + 2\log(1/\delta))/\epsilon^2_m. \]
\begin{proof}
The detailed proof is provided in Appendix~F.
\end{proof}
\begin{remark}
This relation helps quantify the required magnitude of DP noise with key parameters to maintain the desired privacy guarantee. Notably, \(\sigma_m^2\) exhibits a negative correlation with the privacy budget \(\epsilon_m\): as \(\epsilon_m\) increases, the acceptable privacy leakage tolerance grows, thereby reducing the required noise variance. Conversely, \(\sigma_m^2\) is quadratically and positively correlated with the client sampling rate \(q_m\) as a higher sampling ratio increases a client's participation frequency, thereby elevating the risk of privacy leakage and necessitating stronger noise injection. The noise level also grows linearly with the number of rounds \(T\), reflecting the cumulative privacy loss over time. In practice, one may choose the exact lower bound value that minimizes the magnitude of DP noise.
\end{remark}
\end{theorem}

In addition to the per-group privacy guarantees provided by GDPFed, we also establish its overall (system-level) privacy guarantee. To this end, we first present the principle of parallel composition for DP mechanisms, as stated in \autoref{lemma:parallelCom}. 

\begin{lemma}[Parallel Composition of DP~\cite{yu2019differentially}]\label{lemma:parallelCom} 
Let $\{\mathcal{D}_m\}_{m \in [M]}$ be a partition of the input domain $\mathcal{D}$ into disjoint subsets. Suppose each randomized mechanism $\mathcal{M}_m: \mathcal{D}_m \rightarrow \mathbb{R}^d$ satisfies $(\epsilon_m, \delta)$-DP. Then, any (possibly randomized) function applied to the collection $\{\mathcal{M}_m\}_{m \in [M]}$ satisfies $(\max_{m \in [M]} \epsilon_m, \delta)$-DP.
\end{lemma}
Intuitively, when the input domain is partitioned into disjoint subsets independently of the actual data, and a DP mechanism protects each subset, the weakest mechanism determines the overall privacy guarantee, that is, the one with the largest privacy budget $\epsilon_m$. Using \autoref{lemma:parallelCom}, we can establish the system-level privacy guarantee of GDPFed in \coroautoref{coro:privacy_guarantee_of_GDP}.

\begin{corollary}[Privacy Guarantee of GDPFed]\label{coro:privacy_guarantee_of_GDP} 
\revision{Assume that the local datasets of all clients are pairwise disjoint, i.e., no data point is shared between any two clients. If, in GDPFed, each group $m \in [M]$ selects a noise multiplier $\sigma^2_m$ satisfying the condition in \autoref{thm:privacy_loss_per_group}, then after $T$ training rounds, GDPFed provides
$\{(\epsilon_m, \delta)\}_{m \in [M]}$ group-wise DP guarantees,
and the entire system satisfies $(\max_{m \in [M]} \epsilon_m, \delta)$-DP by the parallel composition in \autoref{lemma:parallelCom}.}
\end{corollary}

\begin{remark}
Intuitively, we observe that 
$
\min_{i \in [n]} \epsilon_i\leq \max_{m \in [M]} \min_{i \in \mathcal{G}_m} \epsilon_i \leq \max_{i \in [n]} \epsilon_i,
$
where both inequalities become equalities under the homogeneous DP setting (i.e., $\epsilon_i = \epsilon$ for all clients).  Compared with DP-FedAvg, which guarantees $(\min_{i \in [n]} \epsilon_i, \delta)$-DP for every client, GDPFed yields a slightly weaker overall guarantee. Nevertheless, it flexibly accommodates heterogeneous privacy requirements by allowing groups with looser privacy budgets to use smaller noise multipliers, which in turn can improve the utility of the model.
\revision{
In practice, datasets owned by different groups may contain overlapping data points, which violates the disjointness condition required by \autoref{lemma:parallelCom}. 
Consider the case where groups $\mathcal{G}_i$ and $\mathcal{G}_j$ have overlapping datasets and an individual record appears in both groups. 
In this case, removing this record affects both DP mechanisms $\mathcal{M}_i$ and $\mathcal{M}_j$, and thus changes their joint output. The ratio between the output distributions on adjacent datasets is given by the product of the likelihood ratios of $\mathcal{M}_i$ and $\mathcal{M}_j$, which leads to an effective privacy budget that is upper bounded by $\epsilon_i + \epsilon_j$ (sequential composition), rather than $\max \{ \epsilon_i, \epsilon_j \} $ (parallel composition). 
As a result, the overall privacy guarantee becomes
$
\left(
\max\left\{
\epsilon_i + \epsilon_j,\;
\max_{m \in [M]\setminus\{i,j\}} \epsilon_m
\right\},
\delta
\right)\text{-DP},
$
which is weaker than or equal to the ideal disjoint case, since
$
\max\left\{
\epsilon_i + \epsilon_j,\;
\max_{m \in [M]\setminus\{i,j\}} \epsilon_m
\right\}
\ge
\max_{m \in [M]} \epsilon_m.
$
Since our goal in this work is to characterize the system-level privacy guarantee of the GDPFed system theoretically, we assume that each client’s dataset is disjoint from all others to ensure that the parallel composition result in \autoref{lemma:parallelCom} applies. In our empirical evaluation, we also enforce this assumption by explicitly constructing non-overlapping client datasets.
}
\end{remark}

\subsection{Analyzing DP Noise} 
Building upon the privacy analysis of GDPFed, we now conduct a detailed investigation of the factors that influence the magnitude of the DP noise applied to the model updates, aiming to derive further insights for improving model utility.
In GDPFed, we leverage the Gaussian mechanism to impose noise for each group, drawn from the distribution \(\mathcal{N}(0, (C^2\sigma^2_m/r_m) \cdot~\mathbf{I}^d)\), thereby ensuring $(\epsilon_m, \delta)$-DP. The expected squared $\ell_2$-norm of the total noise applied to aggregated model updates (denoted as $\Lambda_m$) received by the server is $\Lambda_m = d\cdot C^2\sigma_m^2$, for group $m$. Substituting \(\sigma_m^2\) with its lower bound from \autoref{thm:privacy_loss_per_group}, we obtain \( \Lambda_m = 7dq_m^2T(\epsilon_m + 2\log(1/\delta))C^2 /\epsilon^2_m\). We focus on analyzing the influence of two critical parameters, \(d\) and \(q_m\), on the magnitude of DP noise, as other parameters are typically fixed in a given HDPFL system. Specifically, properly adjusting $d$ and $q_m$ can effectively reduce the amount of noise under the same privacy guarantee. If model utility is preserved in the process, this can potentially lead to improved overall performance.

\textit{a) reducing $d$.} Modern neural network architectures (e.g., ResNet~\cite{resnet}) are typically designed with millions of parameters to ensure strong generalization capability. This results in a large model dimensionality \(d\), which in turn significantly increases the magnitude of DP noise. To reduce $d$, existing works consider low-rank decomposition~\cite{lora3, lora1, lora2} or structured pruning~\cite{sp1, sp2}. However, these methods suffer from significant utility loss~\cite{fedsmp}. Moreover, they alter the model architecture, which poses challenges for model aggregation in FL.
A more effective approach is to retain the original architecture while reducing the number of active parameters, a technique known as \textit{model sparsification}~\cite{spar_1} (i.e., unstructured pruning). This strategy selectively eliminates a subset of model parameters, which directly reduces DP noise while preserving both the original network architecture and model performance, leveraging the natural redundancy present in deep neural networks.

\textit{b) Adjusting $q_m$.} Regarding \(q_m\), directly reducing it leads to a smaller magnitude of DP noise injected into model updates for group \(m\). Intuitively, it is desirable to reduce the sampling probability for groups with tighter privacy requirements (i.e., smaller \(\epsilon_m\)), as these groups demand lower privacy loss. In practice, privacy-sensitive clients indeed prefer to participate less frequently in training, reducing their exposure to potential inference attacks~\cite{ccs_rdp}. However, this approach degrades global model performance if insufficient clients participate in local training. Assuming a minimum global participation ratio $q$ is required (i.e., in expectation, $qn$ clients are selected for local training in each round of GDPFed), clients with larger privacy budgets should participate more frequently, as their model updates contain less noise. Yet, excessive participation frequency also increases DP noise under the same privacy guarantee. Consequently, there exists an \textit{optimal set of client sampling ratios} that balances these competing factors while satisfying both participation and privacy constraints.

\begin{algorithm}[t]
\small
\caption{GDPFed$^+$: Sparsification-Amplified GDPFed with Optimal Client Sampling}
\label{alg:GDPFed}
\begin{algorithmic}[1]
\REQUIRE Optimal client sampling ratio $\{q_m\}^M_{m=1}$; training rounds $T$; local iteration $\tau$; local learning rate $\eta$; clipping threshold $C$; noise multipliers $\{\sigma_m^2 \}^M_{m=1}$; reweighting parameters $\{\omega_m \}^M_{m=1}$; top-$k$ parameter $\{k_m \}^M_{m=1}$;
\ENSURE Global model $\mathbf{\theta}^{T}$

\STATE \textbf{Initialization:} Randomly initialize $\theta^{0} \in \mathbb{R}^d$

\FOR{$t = 0$ to $T{-}1$}
    \FOR{group $m = 1$ to $M$}
        \STATE Sample $r_m=q_m|\mathcal{G}_m|$ clients $\mathcal{S}^t_m $ from $\mathcal{G}_m$
        
        \STATE Broadcast $\theta^t$ to all clients in $\mathcal{S}^t_m$ \label{alg:GDPFed_broadcase_mask}
        
        \FOR{client $i \in \mathcal{S}^t_m$ \textbf{in parallel}}
            \FOR{$s = 0$ to $\tau{-}1$}
                \STATE Compute a mini-batch gradient $g_{m, i}^{t,s}$

                \STATE ${\theta}_{m, i}^{t, s+1} \leftarrow \theta_{m, i}^{t, s} - \eta   g_{m, i}^{t,s}$
            \ENDFOR
            \STATE $\hat{\Delta}_{m, i}^t \leftarrow {\theta}_{m, i}^{t,\tau} - \theta^{t}$

            \STATE $\bar{\Delta}_{m, i}^t \leftarrow \hat{\Delta}_{m, i}^t \times \min(1, C/\|\hat{\Delta}_{m, i}^t\|_2)$

            \STATE $\Delta_{m, i}^t \leftarrow \bar{\Delta}_{m, i}^t + \mathcal{N}(0, (C^2\sigma^2_m/r_m) \cdot \mathbf{I}^d)$ \label{alg:GDPFed_addnoise}
            

            \STATE $\mathbf{y}^t_{m,i} \leftarrow \mathrm{Encrypt}(\Delta_{m,i}^t)$ via secure aggregation and send $\mathbf{y}^t_{m, i}$ to the server \label{alg:secure_agg}
        \ENDFOR
        \STATE $\bar{\mathbf{y}}^t_{m} \leftarrow \sum_{i \in \mathcal{S}^t_m} \mathbf{y}^t_{m, i}$ \label{alg:sec}

        \STATE  $\tilde{\mathbf{y}}^t_{m} \leftarrow \mathrm{Top}_k(\bar{\mathbf{y}}^t_{m}, k_m)$ \label{alg:GDPFed_calculate_per_group_mask}
    \ENDFOR

    \STATE $\theta^{t+1} \leftarrow \theta^t + \sum_{m\in[M]} \omega_m \tilde{\mathbf{y}}^t_{m}$ \label{alg:refinement}
\ENDFOR
\RETURN $\theta^T$
\end{algorithmic}
\end{algorithm}

\begin{algorithm}[t]
\small
\caption{Top-$k$ Sparsifier $\mathrm{Top}_k(\cdot)$}
\label{alg:topk_mask}
\begin{algorithmic}[1]
\REQUIRE Vector $x \in \mathbb{R}^d$, top-$k$ parameter $k \in [0,d]$
\ENSURE Binary mask $\mathbf{mk} \in \{0,1\}^d$

\STATE \textbf{Initialization:} Initialize $\mathbf{mk} \gets \mathbf{0}^d$
\STATE Compute absolute values: $a_j \gets |x_j|$ for all $j \in [1,d]$
\STATE Sort indices $\pi$ such that $a_{\pi(1)} \geq a_{\pi(2)} \geq \dots \geq a_{\pi(d)}$
\FOR{$j = 1$ \TO $k$}
    \STATE $[\mathbf{mk}]_{\pi(j)} \gets 1$
\ENDFOR
\RETURN $\mathbf{mk} \odot x$
\end{algorithmic}
\end{algorithm}

\section{Sparsification-Amplified GDPFed with Optimal Client Sampling} \label{sec: gdpfedplus}

In this section, we introduce an enhanced variant of GDPFed, termed GDPFed$^{+}$, which improves model utility by incorporating sparsification techniques that reduce the model dimension $d$ and by deriving optimal client sampling ratios for each group.
The algorithm of GDPFed$^+$ is detailed in \algautoref{alg:GDPFed}.


\subsection{GDPFed with Per-group Sparsification}
To achieve client-level DP under our attack model, sampled clients in each group add a small amount of noise to the model updates (\lineautoref{alg:GDPFed_addnoise} in \algautoref{alg:GDPFed}) and send them to the server via secure aggregation (\lineautoref{alg:secure_agg}). The secure aggregation ensures the server only receives the sum of model updates from each group, as well as the summed noise (\lineautoref{alg:sec}). Here, per-group DP perturbation is applied over the entire parameter space (i.e., $\mathbb{R}^d$) of model updates. In other words, all parameters are subjected to perturbation regardless of their importance. However, prior studies have shown that neural networks typically exhibit substantial parameter redundancy, with many parameters contributing negligibly to the task~\cite{spar_1, spar_2, spar_3}. Under DP settings, perturbing unimportant parameters introduces redundant noise, unnecessarily degrading utility. 

A practical remedy is model sparsification, which removes unimportant parameters from model updates along with their associated noise. Specifically, a top-$k$ sparsifier, denoted as $\mathrm{Top}_k(\cdot)$, is applied to retain only the $k \in [0, d]$ most important parameters. Note that $k = 0$ corresponds to eliminating all parameters, while $k = d$ indicates no sparsification. The detailed algorithm of $\mathrm{Top}_k(\cdot)$ is provided in \algautoref{alg:topk_mask}. In this work, we adopt a widely-used and straightforward criterion for identifying important parameters—their absolute magnitude~\cite{spar_2, fedsmp, alignins}. It should be noted that sparsification must be applied after DP perturbation to preserve the desired $(\epsilon, \delta)$-DP guarantee, as ensured by the post-processing property of DP given as in \autoref{lemma:post-processing}.

\begin{lemma}[Post-Processing of DP~\cite{dp_og}]\label{lemma:post-processing} Let $\mathcal{M}$ be a randomized mechanism that satisfies $(\epsilon, \delta)$-DP. Then, for any mapping $g$, the composed function $g \circ \mathcal{M}$ also satisfies $(\epsilon, \delta)$-DP. 
\end{lemma}

Technically, in training round~$t$, the server applies the $\mathrm{Top}_k(\cdot)$ sparsifier to the per-group model updates summation $\bar{\mathbf{y}}^t_m$ using a group-specific sparsification parameter $k_m$, resulting in sparsified updates $\tilde{\mathbf{y}}^t_m$ (\lineautoref{alg:GDPFed_calculate_per_group_mask}). These sparsified updates are then aggregated with reweighting parameters to refine the global model (\lineautoref{alg:refinement}). In the following, we theoretically analyze the error introduced by sparsification.

\subsection{Bounded Sparsification Error} To reflect varying privacy preferences, it is desirable to assign distinct sparsification parameters (i.e., $k_1, k_2, \dots, k_M$) to different groups. Intuitively, groups with stricter privacy requirements should be assigned more aggressive sparsification to mitigate the larger DP noise added to their updates. However, \(\mathrm{Top}_k(\cdot)\) is not without cost since using a smaller \(k_m\) means that more parameters are removed, which can potentially lead to a non-negligible loss in utility. To formally quantify this relationship, we introduce \autoref{lemma: bounded sparsifier}, which characterizes the approximation error introduced by the \(\mathrm{Top}_k(\cdot)\) sparsifier. 

\begin{lemma}[Bounded Sparsification]
\label{lemma: bounded sparsifier}
Given a vector $x\in\mathbb{R}^d$ and a sparsification parameter $k \in [0, d]$. We have $\mathbb{E}\| \mathrm{Top}_k(x)-x\|^2 \leq \phi\|x\|^2$, where $\phi$ is a sparsification error coefficient.
\end{lemma}

It is evident that a smaller \(k\) results in a larger \(\phi\), thereby leading to a greater sparsification error. Therefore, $k_m$ should be carefully selected in order to successfully leverage its benefit. In the literature, $\phi$ is typically set to $1-k/d$~\cite{fedsmp, alignins} or $(1-k/d)^2$~\cite{shi2019understanding} to measure the sparsification error. In this work, we choose $\phi = (1-k/d)^2$ as it provides a tighter bound.

\subsection{Convergence Analysis of GDPFed}
We first present several important assumptions that help us conduct the convergence analysis. 
\begin{assumption}[$L$-Smoothness] \label{assp:lsmooth}
The local objective $f_{m,i}(\cdot)$ of each client $i \in \mathcal{G}_m$ in any group $m \in [M]$, is $L$-smooth with constant $L > 0$; i.e., for all $x, y \in \mathbb{R}^d$,
$\left\| \nabla f_{m,i}(x) - \nabla f_{m,i}(y) \right\| \leq L \left\| x - y \right\|$.
\end{assumption}

\begin{assumption}[Unbiased Gradient and Bounded Variance] \label{assp:bounded_divergence}
For each client $i \in \mathcal{G}_m$ in any group $m \in [M]$, the stochastic gradient $g_{m,i}(x) \in \mathbb{R}^d$ satisfies:
$\mathbb{E}[g_{m,i}(x)] = \nabla f_{m,i}(x)$ and $\mathbb{E} \left\| [g_{m,i}(x)]_j - [\nabla f_{m,i}(x)]_j \right\|^2 \leq \zeta_{m,i}^2, \forall j\in[d],$
where the expectation is over mini-batch sampling. 
\end{assumption}

\begin{assumption}[Bounded Dissimilarity] \label{assp:bounded_dissimilarity}
    There exist $\beta^2 \geq 1$ and $\kappa^2 \geq 0$ such that $\sum_{m=1}^{M} \omega_m \sum_{i \in \mathcal{G}_m} \left\| \nabla f_{m,i}(x) \right\|^2 
        \leq \beta^2 \| \sum_{m=1}^{M} \omega_m \sum_{i \in \mathcal{G}_m} \nabla f_{m,i}(x) \|^2 + \kappa^2.$
     With identical local objectives, the inequality holds with $\beta^2 = 1$ and $\kappa^2 = 0$.

\end{assumption}

Note that \autoref{assp:lsmooth}--\ref{assp:bounded_divergence} are commonly used in the theoretical analysis of distributed learning systems~\cite{lasa, nesterov2018lectures, fedsmp}. In particular, \autoref{assp:bounded_divergence} bounds the coordinate-wise variance of local gradients~\cite{hu2021federated}. Meanwhile, \autoref{assp:bounded_dissimilarity} captures  inter-client heterogeneity in FL~\cite{x1, x2, x3}.
With the above assumptions, we provide the convergence result of GDPFed under the general non-convex setting in \autoref{the:convergence_analysis}.

\begin{theorem}[Convergence Result of GDPFed] \label{the:convergence_analysis} Let $\theta^0$ be the initial point and $f^*$ be the optimal objective value. Assume the learning rate satisfies $\eta \leq \min\{1/\left(4L\beta^2\left( \tau  +  1 \right) + 8L \tau \beta^2\right), 1/(16\tau L)\}$, then the sequence of outputs $\theta^t$ generated by GDPFed satisfies:
\begin{equation*} \label{formula:convergence_bound}
\begin{aligned}
    \frac{1}{T} \sum_{t=0}^{T-1} \left\| \nabla f(\theta^t) \right\|^2 
    &\leq \frac{8 \left(f(\theta^0) - f^*\right)}{\eta T \tau} + \mu_1 \kappa^2 \\ & + \mu_2 \sum_{m=1}^{M} \omega_m (\phi_m + 1)d \zeta_m^2  \\ & + \mu_3 \sum_{m=1}^{M} \frac{k_m \omega_m^2 C^2 \sigma_m^2}{r_m q_m},
\end{aligned}
\end{equation*}
where $\mu_1=4L \eta \tau + 4L  \eta + 64 L $, $ \mu_2=32 L \eta \tau +  L  \eta  +  L \eta /\tau$, $\mu_3=4L/\eta\tau$, $\phi_m = (1-k_m/d)^2$, and $\zeta_m^2 = (1/|\mathcal{G}_m|) \sum_{i \in \mathcal{G}_m} \zeta_{m,i}^2$.
\begin{proof}
    The detailed proof is given in Appendix~G.
\end{proof}
\begin{remark}
\label{remark: convergence}
If \(\phi_m = 0, \ \forall m\in[M]\), meaning no sparsification is applied, the first three terms on the right-hand side of the convergence bound correspond to the optimization error of FedAvg. In particular, the third term captures group-wise heterogeneity in model updates, which are influenced by the group-wise sparsification parameters $k_m$. Specifically, applying more aggressive sparsification (i.e., smaller $k_m$) increases the heterogeneity among per-group model updates. However, as reflected in the final term of the bound, a smaller $k_m$ reduces the privacy error introduced by DP, confirming our analysis in \autoref{sec: gdpfed}. This highlights a fundamental trade-off: selecting an appropriate $k_m$ is crucial for balancing sparsification and privacy errors, thereby minimizing the overall convergence error. Hence, by directly minimizing the errors in the third and last terms that are related to $k_m$, we obtain a coarse closed-form expression for the optimal sparsification level for the group $m$:
$k^*_m/d = 1 - 2 \omega_m \sigma_m^2 / (\eta \tau \mu_4 r_m^2), \ \forall m \in [M],$
where $\mu_4 = 32 \eta \tau + \eta + \eta / \tau$. At this case, $\phi^*_m$ is given by
$\phi^*_m = 4 \omega_m^2 \sigma_m^4/(\eta \tau \mu_4 r_m^2)^2$ (see the sketch of the derivation in Appendix~H). This yields a tighter upper bound for the convergence error. \revision{Importantly, $\phi_m^*$ can be directly specified from the system configuration.}


\end{remark}
\end{theorem}

\subsection{Optimal Client Sampling Ratios}
Building on the convergence analysis of sparsification-amplified GDPFed, we now discuss how to determine the optimal client sampling ratio for each group. To ensure that the global model trained by GDPFed converges to a better optimum, it is desirable to minimize the true gradient of the objective function (i.e., the left-hand side of the convergence result). However, directly minimizing this function is typically infeasible in practice, as $\nabla f (\theta^t)$ is a high-dimensional, non-convex function. An alternative approach is optimizing its upper bound (i.e., the right-hand side of the convergence bound), which approximates optimizing the objective function. Notably, only the third and last terms in the bound are influenced by the client sampling ratios. This leads to the constrained minimization problem formulated in \problemautoref{problem:opt}.

\begin{problem}[Optimal Sampling Ratios for GDPFed] \label{problem:opt} 
The optimal per-group sampling ratios \( \{q_m\}_{m \in [M]} \) for GDPFed are obtained by solving the following constrained optimization problem:
\begin{equation*}
    \begin{aligned}
        \min_{\{q_m\}_{m \in [M]}} \ & 
        \sum_{m \in [M]}  \omega_m \left(\mu_4 (1 + \phi^*_m) + \mu_5 \frac{\left(1 - \sqrt{\phi^*_m}\right) \omega_m \sigma_m^2}{r_m^2} \right) \\
        \text{s.t.} \quad & r_m = q_m |\mathcal{G}_m|, \ \sum_{m \in [M]} r_m = qn, \ 
    \end{aligned}
\end{equation*}
where \( \mu_4 = 32 \eta \tau + \eta + \eta/\tau \) and \( \mu_5 = 4 / (\eta \tau) \).
\end{problem}

\begin{remark}
The optimal sparsification error coefficient \(\phi^*_m\) is defined as given in \autoref{remark: convergence}. If there is no sparsification applied, then \(\phi^*_m=0, \ \forall m \in [M]\). The noise multiplier \(\sigma_m^2\) required to satisfy the \((\epsilon_m, \delta)\)-DP for group \(m\) is derived in \autoref{thm:privacy_loss_per_group}. All parameters in \problemautoref{problem:opt} are now determined by the system settings, as terms such as $L$ and $\zeta$ have been eliminated (see the formulation sketch in Appendix~I for details), leaving only the decision variables. Therefore, the optimal client sampling ratios for each group in GDPFed can be efficiently obtained by solving this minimization problem. As \problemautoref{problem:opt} is a non-convex optimization problem, one can resort to existing solvers in practice, such as optimization libraries in Python (e.g., scikit-learn~\cite{sklearn}), to obtain a feasible solution.
\end{remark}

With the optimal client sampling ratios derived from solving \problemautoref{problem:opt}, which minimizes the convergence upper bound in \autoref{the:convergence_analysis}, GDPFed$^+$ converges to a better minimum than GDPFed, thereby enhancing model utility. 
Importantly, GDPFed$^+$ still satisfies the per-group privacy guarantees in \autoref{thm:privacy_loss_per_group}, the overall privacy guarantee in \coroautoref{coro:privacy_guarantee_of_GDP}, and the convergence bound in \autoref{the:convergence_analysis}. 
In the following evaluation section, we empirically validate the performance of the proposed GDPFed and GDPFed$^+$ methods, which are equipped with per-group sparsification and optimal client sampling ratios, against several baseline approaches.

\section{Empirical Evaluation}

\label{sec: evaluation}

\subsection{Experimental Settings}

\begin{table}[t]
\centering
\caption{Detailed system configurations for each dataset.}
\label{tab: hyperparameters}
\scalebox{0.87}{
    \begin{tabular}{l|ccccc}
    \toprule
    \textbf{Dataset} & $n$ & $T$ & $C$ & $(\epsilon_1, \epsilon_2, \epsilon_3)$ & $(q_1, q_2, q_3)$-$q$ $(\%)$ \\
    \midrule
    FMNIST & $6{,}000$  & $50$ & $1.5$ & $(0.5, 1.5, 3.0)$ & $(0.69, 1.89, 3.42)$-$2$ \\
    SVHN & $6{,}000$ & $100$ & $1.0$ & $(0.5, 1.5, 3.0)$ & $(1.66, 4.67, 8.68)$-$5$ \\
    Shakespeare & $714$ & $50$ & $1.0$ & $(0.5, 1.5, 3.0)$ & $(4.79, 9.83, 15.21)$-$10$ \\
    CIFAR-10 & $600$ & $100$ & $1.5$ & $(2.0, 6.0, 12.0)$ & $(3.61, 9.62, 16.77)$-$10$ \\
    \bottomrule
    \end{tabular}
}
\end{table}

\subsubsection{\revision{Dataset settings}} \label{sec: dataset_settings}Our evaluation covers four benchmark datasets for HDPFL: Fashion MNIST (FMNIST)~\cite{FMNIST}, SVHN~\cite{SVHN}, CIFAR-10~\cite{cifar10_100}, and Shakespeare~\cite{sha}. Correspondingly, we adopt a $2$-layer CNN for FMNIST, a $3$-layer CNN for SVHN, a ResNet-18~\cite{resnet} for CIFAR-10, and an LSTM model for Shakespeare. 
We conduct experiments in \textit{cross-device} FL settings with $n$ clients. Client local datasets are assumed to be \textit{independent and identically distributed} (IID), except Shakespeare, which is evaluated in its inherent non-IID form. We summarize the system configurations in \autoref{tab: hyperparameters}.
%

\subsubsection{Baselines} \label{sec: baseline_settings} We compare against four baselines to demonstrate the effectiveness of GDPFed$^+$. Specifically, we include two important baselines: \revision{Pure FedAvg (P-FedAvg), a non-private FL case that serves as a strong reference point for model utility without privacy noise}, and client-level DP-FedAvg (DP-FedAvg), which enforces the strictest privacy requirement across all clients. Moreover, our comparisons include state-of-the-art methods IDP-FedAvg~\cite{boenisch2023have} and PFA~\cite{PFA}. We follow their default settings, with further details provided in Appendix~C.

\subsubsection{Training settings} \label{sec: training_settings} In all experiments, local clients use \textit{stochastic gradient descent} (SGD) as the optimizer with a learning rate of $\eta = 0.1$ and a decay ratio of $0.99$. \revision{For the FMNIST, SVHN, CIFAR-10, and Shakespeare datasets, we set the momentum coefficient to $0.0$, $0.0$, $0.5$, and $0.9$, the number of local training iterations $\tau$ to $5$, $25$, $5$, and $30$, and the batch size to $10$, $10$, $50$, and $4$, respectively.} We set the uniform DP failure parameter as $\delta = 1/n^{1.1}$, following the recommendation in~\cite{fedsmp, dp-fedavg}. To reweight the per-group model updates, we set the reweighting parameter $\omega_m = (1/qn) \cdot r_m^2 / \sum_{m \in [M]} r_m^2$ for all $m \in [M]$. This reweighting strategy prioritizes groups with higher expected client participation and helps reduce the total noise added to the aggregated model. Note that to ensure a fair comparison, we adopt this reweighting parameter for all group-based methods, including PFA, IDP-FedAvg, GDPFed, and GDPFed$^+$. Additional baseline settings are provided in Appendix~C. \textit{All experiments are repeated three times with different random seeds to ensure statistical reliability.}

\subsubsection{System settings} \label{sec: system_settings}
Following prior works~\cite{alaggan2015heterogeneous, boenisch2023have, kiani2025differentially} that simulate heterogeneous privacy requirements, clients are assigned to one of \textit{three} groups, each associated with a distinct minimum privacy budget $(\epsilon_1, \epsilon_2, \epsilon_3)$. 
By default, clients are evenly distributed among three groups. 
For GDPFed$^+$, the optimal client sampling ratios $(q_1, q_2, q_3)$ \revision{are} derived by solving \problemautoref{problem:opt}. By default, the sparsification levels $k_m/d$ for each group are $(0.7, 0.8, 0.9)$. 

\subsubsection{Hardware settings} \label{sec: hardware_settings} All experiments were conducted on a Linux-based internal compute cluster equipped with $8$ NVIDIA RTX A$6000$ GPUs (each with \revision{$48$} GB of memory) and AMD EPYC $7763$ $64$-core CPUs.

\subsection{Experimental Results}
\label{sec: main_results}

\begin{figure}[t]
    \centering
    \includegraphics[width=0.9\linewidth]{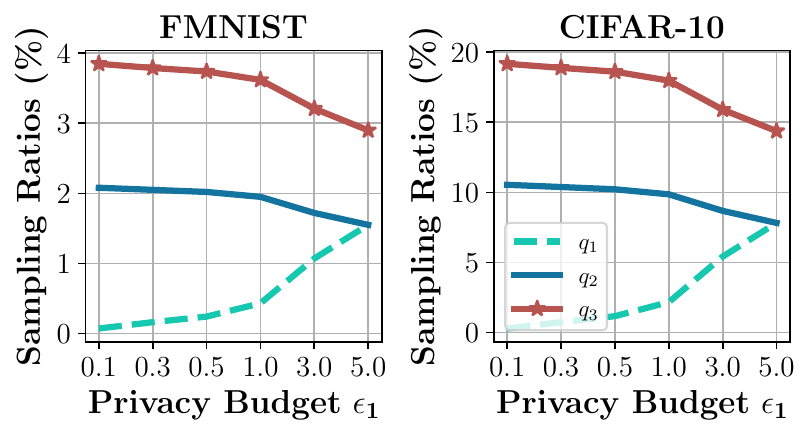}
    \caption{Optimal client sampling ratios under varying $\epsilon_1$. Sampling ratio for each group is adjusted dynamically to satisfy the global constraint with fixed $\epsilon_2$ and $\epsilon_3$.}
    \label{fig:samplingratios}
\end{figure}



\subsubsection{Dynamics of client sampling ratios} We first demonstrate how the optimal client sampling ratios dynamically adjust in response to varying privacy budgets, as dictated by our optimization formulation in \problemautoref{problem:opt}. Specifically, we conduct experiments on the FMNIST and CIFAR-10 datasets. In both cases, we fix the privacy budgets for Group $2$ and Group $3$ as $\epsilon_2 = 1.5$ and $\epsilon_3 = 3.0$ for FMNIST and $\epsilon_2 = 6.0$ and $\epsilon_3 = 12.0$ for CIFAR-10, respectively, and vary $\epsilon_1$ to examine how the optimal sampling ratios evolve. The resulting trends are illustrated in \autoref{fig:samplingratios}. Across both datasets, we observe a consistent pattern governed by the optimization objective. When $\epsilon_1$ is small, the corresponding sampling ratio $q_1$ decreases to accommodate the stronger noise required for stricter privacy. As $\epsilon_1$ increases, $q_1$ rises accordingly, while $q_2$ and $q_3$ adjust downward to maintain the global constraint $\sum_{m \in [M]} r_m = qn$. Notably, when $\epsilon_1 = \epsilon_2 = 1.5$ for FMNIST and $\epsilon_1 = \epsilon_2 = 6.0$ for CIFAR-10, Groups $1$ and $2$ yield identical sampling ratios.

\begin{figure}[t]
    \centering
    \includegraphics[width=0.9\linewidth]{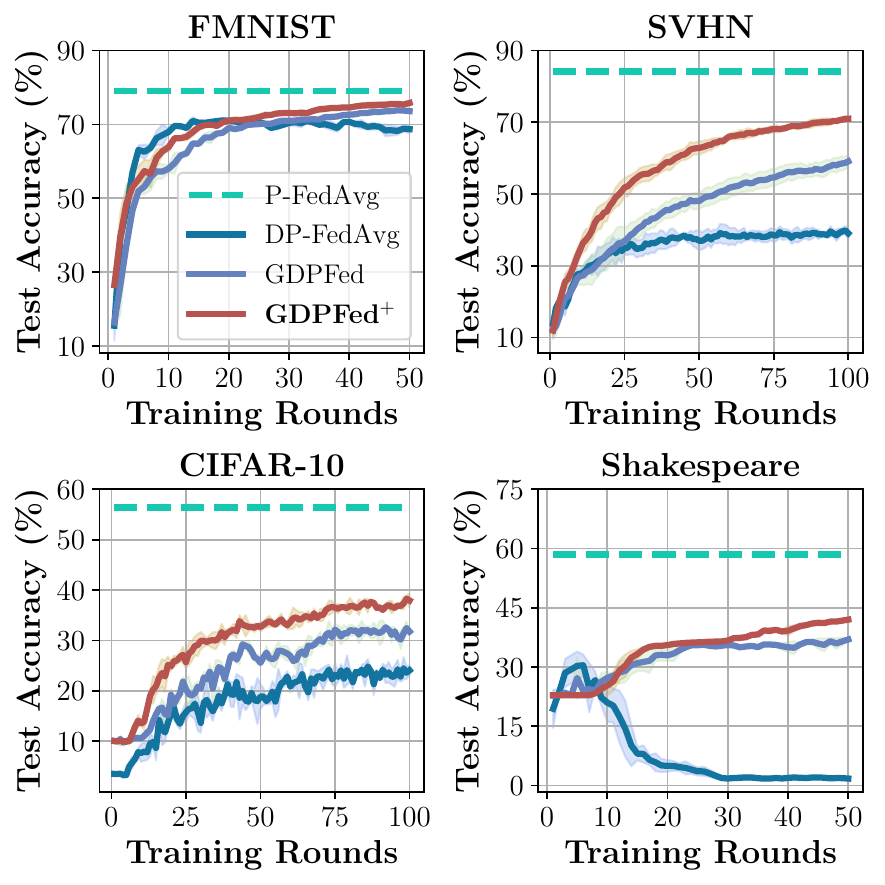}
    \caption{Convergence curve comparison between P-FedAvg, DP-FedAvg, GDPFed, and GDPFed$^+$.}
    \label{fig:convergence}
\end{figure}

\begin{table}[t]
\centering
\caption{Test accuracy ($\%$) of baselines, GDPFed, and GDPFed$^+$ on each dataset.}
\label{tab:main_table}
\scalebox{0.81}{
\begin{tabular}{l|cccc|c}
\toprule
\textbf{Method} & \textbf{FMNIST} & \textbf{SVHN} & \textbf{CIFAR-10} & \textbf{Shakespeare} & \textbf{Avg.} \\
\midrule
P-FedAvg & $78.96_{\pm0.90}$ & $84.13_{\pm0.18}$ & $56.40_{\pm0.34}$ & $60.62_{\pm0.76}$ & $70.03$ \\
\midrule
\midrule
DP-FedAvg & $71.88_{\pm0.15}$ & $40.80_{\pm1.73}$ & $32.10_{\pm 0.34}$ & $34.97_{\pm0.76}$ & $44.94$ \\
GDPFed & $73.97_{\pm0.21}$ & $59.11_{\pm1.72}$ & $34.00_{\pm0.44}$ & $37.63_{\pm0.29}$ & $51.18$ \\
PFA & $73.67_{\pm 0.28}$ & $67.51_{\pm0.84}$ & $37.17_{\pm 0.98}$ & $36.69_{\pm0.21}$ & $53.76$ \\
IDP-FedAvg & $74.80_{\pm 0.19}$ & $66.46_{\pm0.98}$ & $37.49_{\pm 0.76}$ & $37.26_{\pm0.55}$ & $54.00$ \\
\rowcolor[rgb]{ .9, .9, .9} \textbf{GDPFed$^+$} & $\mathbf{75.83_{\pm0.47}}$ & $\mathbf{71.10_{\pm0.58}}$ & $\mathbf{38.78_{\pm0.36}}$ & $\mathbf{42.00_{\pm0.45}}$ & $\mathbf{56.93}$ \\
\bottomrule
\end{tabular}
}
\end{table}

\subsubsection{Convergence and main results} \label{sec: eva_convergence}
We present the convergence curves of P-FedAvg, DP-FedAvg, GDPFed, and GDPFed$^+$ in \autoref{fig:convergence}, together with the test accuracies in \autoref{tab:main_table}. \revision{As shown in \autoref{fig:convergence}, DP-FedAvg exhibits clear performance degradation compared to P-FedAvg. This is because DP-FedAvg enforces the strictest privacy budget uniformly across clients, resulting in strong noise injection at every training round. On the Shakespeare task, DP-FedAvg initially achieves a modest accuracy improvement, but its performance subsequently degrades over time. 
In the early training rounds, the cumulative DP noise remains limited, allowing the model to capture useful patterns. However, as training progresses, the accumulated noise grows across rounds, and due to the high sensitivity of the Shakespeare task to DP noise, it gradually overwhelms the learning signal of the model, ultimately causing model collapse.
}

In contrast, GDPFed enforces DP at the group level rather than using a single global privacy budget. By reducing unnecessary noise for clients with looser privacy requirements, GDPFed consistently converges to better solutions across all datasets.
Building upon this, GDPFed$^+$ further integrates model sparsification with optimal per-group client sampling ratios, resulting in enhanced performance and achieving the highest test accuracies on all datasets. Moreover, GDPFed$^+$ exhibits a more stable convergence process compared to other baselines. As shown in \autoref{tab:main_table}, GDPFed$^+$ achieves an average accuracy of \(56.93\%\) across all datasets, representing a \(+5.75\%\) improvement over GDPFed’s accuracy of \(51.18\%\). Moreover, GDPFed$^+$ surpasses two state-of-the-art baselines, PFA and IDP-FedAvg, with average accuracy gains of \(+3.17\%\) and \(+2.93\%\), respectively. These results empirically demonstrate the effectiveness of GDPFed$^+$ in enhancing model utility while maintaining client-level HDP guarantees.

\begin{table}[t]
\centering
\caption{Noise multipliers ($\sigma^2 $ or $ (\sigma^2_1, \sigma^2_2, \sigma^2_3)$) and noise magnitude ($\Lambda$) on FMNIST and CIFAR-10 across methods.}
\label{tab:complete_siga}
\scalebox{0.95}{
\begin{tabular}{l|cc}
\toprule
\textbf{Method} & \textbf{FMNIST} & \textbf{CIFAR-10} \\
\midrule
DP-FedAvg
& $2.26$ / $5.09$
& $3.52$ / $7.91$ \\
GDPFed
& $(2.26, 0.90, 0.53)$ / $2.77$
& $(3.52, 0.95, 0.49)$ / $3.72$ \\
GDPFed-$\mathrm{opc}$
& $(1.42, 0.87, 0.70)$ / $0.57$
& $(0.98, 0.91, 0.83)$ / $0.64$ \\
\rowcolor[rgb]{ .9, .9, .9}
\textbf{GDPFed$^+$}
& $(\mathbf{1.42}, \mathbf{0.87}, \mathbf{0.70})$ / $\mathbf{0.49}$
& $(\mathbf{0.98}, \mathbf{0.91}, \mathbf{0.83})$ / $\mathbf{0.56}$ \\
\bottomrule
\end{tabular}
}
\end{table}
\subsubsection{Noise amount analysis} Here, we provide a detailed study explaining why GDPFed achieves better performance than DP-FedAvg, and how GDPFed$^+$ further enhances the model utility of GDPFed. Specifically, we compute the total amount of noise (measured as the expectation of the squared $\ell_2$-norm and denoted by $\Lambda$) added to the global model updates by DP-FedAvg, GDPFed, and GDPFed$^+$. In \autoref{tab:complete_siga}, we report the noise multipliers and corresponding $\Lambda$ values for DP-FedAvg, GDPFed, GDPFed-$\mathrm{opc}$ (GDPFed with only optimized client sampling ratios), and GDPFed$^+$ on the FMNIST and CIFAR-10 datasets. As shown in \autoref{tab:complete_siga}, GDPFed reduces the total noise by nearly half compared to DP-FedAvg, as it relaxes the privacy constraints for clients with looser requirements. GDPFed-$\mathrm{opc}$ further significantly decreases $\Lambda$ by adjusting the noise multipliers based on optimized client sampling ratios. Finally, GDPFed$^+$ achieves the smallest $\Lambda$ by \textit{additionally} applying model sparsification to eliminate noise associated with less informative model parameters. These results highlight the effectiveness of our design in reducing the amount of DP noise, thereby improving the privacy-utility trade-off.


\subsection{Additional Results with Various Settings}
\label{sec: more_results}

\begin{table*}[t]
    \caption{Impact of client privacy-preference distribution on test accuracy ($\%$).}
    \centering
    \scalebox{1}{
    \begin{tabular}{l|l|cccc|c}
    \toprule
    \textbf{Dataset} & \textbf{Method} & $1:4:1$ & $2:2:2$ & $3:2:1$ & $1:2:3$ & \textbf{Average} \\
    \midrule
    \multirow{2}{*}{FMNIST}
        & DP-FedAvg  & $\textcolor{black}{71.59}_{\pm 0.46}$ & $\textcolor{black}{71.88}_{\pm 0.15}$ & $\textcolor{black}{71.75}_{\pm 0.58}$ & $\textcolor{black}{71.89}_{\pm 0.51}$ & $71.78$ \\
        & \cellcolor[rgb]{ .9,  .9,  .9}\textbf{GDPFed$^+$}     & \cellcolor[rgb]{ .9,  .9,  .9}$\mathbf{\textcolor{black}{76.06}_{\pm 0.38}}$ & \cellcolor[rgb]{ .9,  .9,  .9}$\mathbf{\textcolor{black}{75.83}_{\pm 0.47}}$ & \cellcolor[rgb]{ .9,  .9,  .9}$\mathbf{\textcolor{black}{74.50}_{\pm 0.34}}$ & \cellcolor[rgb]{ .9,  .9,  .9}$\mathbf{\textcolor{black}{76.77}_{\pm 0.42}}$ & \cellcolor[rgb]{ .9,  .9,  .9}$\mathbf{75.79}$ \\
    \midrule
    \multirow{2}{*}{SVHN}
        & DP-FedAvg  & $\textcolor{black}{40.76}_{\pm 1.58}$ & $\textcolor{black}{40.80}_{\pm 1.73}$ & $\textcolor{black}{40.10}_{\pm 1.59}$ & $\textcolor{black}{40.73}_{\pm 1.43}$ & $40.60$ \\
        & \cellcolor[rgb]{ .9,  .9,  .9}\textbf{GDPFed$^+$}     & \cellcolor[rgb]{ .9,  .9,  .9}$\mathbf{\textcolor{black}{72.02}_{\pm 0.74}}$ & \cellcolor[rgb]{ .9,  .9,  .9}$\mathbf{\textcolor{black}{71.10}_{\pm 0.58}}$ & \cellcolor[rgb]{ .9,  .9,  .9}$\mathbf{\textcolor{black}{65.81}_{\pm 1.17}}$ & \cellcolor[rgb]{ .9,  .9,  .9}$\mathbf{\textcolor{black}{75.02}_{\pm 0.60}}$ \cellcolor[rgb]{ .9,  .9,  .9}& \cellcolor[rgb]{ .9,  .9,  .9}$\mathbf{71.49}$ \\
    \midrule
    \multirow{2}{*}{CIFAR-10}
        & DP-FedAvg  & $\textcolor{black}{33.47}_{\pm 0.87}$ & $\textcolor{black}{32.10}_{\pm 0.27}$ & $\textcolor{black}{33.53}_{\pm 0.62}$ & $\textcolor{black}{33.24}_{\pm 0.72}$ & $33.09$ \\
        & \cellcolor[rgb]{ .9,  .9,  .9}\textbf{GDPFed$^+$}     & \cellcolor[rgb]{ .9,  .9,  .9}$\mathbf{\textcolor{black}{38.90}_{\pm 0.19}}$ & \cellcolor[rgb]{ .9,  .9,  .9}$\mathbf{\textcolor{black}{38.78}_{\pm 0.36}}$ & \cellcolor[rgb]{ .9,  .9,  .9}$\mathbf{\textcolor{black}{35.95}_{\pm 1.14}}$ & \cellcolor[rgb]{ .9,  .9,  .9}$\mathbf{\textcolor{black}{40.64}_{\pm 0.33}}$ & \cellcolor[rgb]{ .9,  .9,  .9}$\mathbf{38.57}$ \\
    \midrule
    \multirow{2}{*}{Shakespeare}
        & DP-FedAvg  & $\textcolor{black}{31.48}_{\pm 3.29}$ & $\textcolor{black}{31.48}_{\pm 2.85}$ & $\textcolor{black}{31.11}_{\pm 3.43}$ & $\textcolor{black}{30.99}_{\pm 3.14}$ & $31.27$ \\
        & \cellcolor[rgb]{ .9,  .9,  .9}\textbf{GDPFed$^+$}     & \cellcolor[rgb]{ .9,  .9,  .9}$\mathbf{\textcolor{black}{42.36}_{\pm 0.40}}$ & \cellcolor[rgb]{ .9,  .9,  .9}$\mathbf{\textcolor{black}{42.00}_{\pm 0.45}}$ & \cellcolor[rgb]{ .9,  .9,  .9}$\mathbf{\textcolor{black}{39.47}_{\pm 0.60}}$ & \cellcolor[rgb]{ .9,  .9,  .9}$\mathbf{\textcolor{black}{43.61}_{\pm 0.43}}$ & \cellcolor[rgb]{ .9,  .9,  .9}$\mathbf{41.86}$ \\
    \bottomrule
    \end{tabular}
    }
    \label{tab:client_ratio_accuracy}
\end{table*}

\subsubsection{Impact of privacy preference distribution} \label{sec: distribution} In our default setting, we assume a uniform client distribution across groups such that $|\mathcal{G}_1| = |\mathcal{G}_2| = |\mathcal{G}_3|$ with $\epsilon_1<\epsilon_2<\epsilon_3$. To further evaluate the robustness and effectiveness of our method, we examine three alternative privacy preference distributions. Specifically, these scenarios vary the proportion of clients in $(\mathcal{G}_1, \mathcal{G}_2, \mathcal{G}_3)$ as follows: (1) $1:4:1$--the moderate privacy group $\mathcal{G}_2$ comprises $4/6$ of clients, while the strictest $\mathcal{G}_1$ and loosest $\mathcal{G}_3$ groups each contain $1/6$; (2) $3:2:1$--the strictest privacy group $\mathcal{G}_1$ holds the largest share with $3/6$ of clients, $\mathcal{G}_2$ contains $2/6$ and $\mathcal{G}_3$ contains $1/6$; 
(3) $1:2:3$--the loosest privacy group $\mathcal{G}_3$  comprises $3/6$ of clients, $\mathcal{G}_1$ contains $1/6$ and \revision{$\mathcal{G}_2$}  includes $2/6$. These different distributions reflect realistic deployment scenarios where privacy needs are not evenly distributed. 

The corresponding results are shown in \autoref{tab:client_ratio_accuracy}. Overall, across four datasets and all distributional settings, GDPFed$^+$ consistently outperforms DP-FedAvg, demonstrating robust performance and adaptability to diverse privacy-preference distributions, and underscoring its practicality for real-world FL systems. It is worth noting that GDPFed$^+$ experiences a slight performance drop under the $3:2:1$ setting, compared with other settings. It is reasonable because GDPFed is designed to reduce the privacy budget waste, and in the $3:2:1$ setting, such waste is inherently not significant. 
From another perspective, the performance decline is primarily due to the increased number of clients from the strictest group, which requires adding more noise. For example, on CIFAR-$10$, GDPFed$^+$ samples $14$ clients from the strict group under the $3:2:1$ distribution, compared to only $7$ under the balanced $2:2:2$ setting. This behavior is driven by the influence of $r_m$ in the optimization objective of \problemautoref{problem:opt}. Nevertheless, even in this challenging case, GDPFed$^+$ still achieves substantial improvements over DP-FedAvg, further demonstrating its effectiveness.

\subsubsection{More results on clients with heterogeneous data} \label{sec: eva_hetero_data} \revision{In practical cross-device FL systems, client data distributions are often highly heterogeneous (non-IID) due to different reasons, such as user behavior and sensing environments. Such data heterogeneity leads to large divergence across local model updates, which increases the risk of global model divergence. The application of DP can further exacerbate this issue. Therefore, it is important to evaluate the performance of different methods under varying degrees of data heterogeneity.} We conduct additional experiments to evaluate the performance of our method under the HDPFL system with various heterogeneous data settings. Specifically, we consider the CIFAR-10 dataset and use the \textit{Dirichlet distribution}~\cite{minka2000estimating} to simulate non-IID data across clients, controlled by the non-IIDness parameter $\lambda$. A larger $\lambda$ corresponds to lower data heterogeneity, and vice versa. We experiment with $\lambda = 0.9, 0.7, 0.5$, and a more extreme case of $\lambda = 0.3$. The results are presented in \autoref{tab:heterogeneity_accuracy_avg}. As $\lambda$ increases (i.e., the data becomes more IID), the accuracy of all methods improves accordingly. Notably, GDPFed$^+$ consistently outperforms both DP-FedAvg and IDP-FedAvg, achieving an average improvement of $+2.26\%$ across all heterogeneity levels. These results demonstrate the effectiveness of GDPFed$^+$ in enhancing model utility even under heterogeneous data distributions in client-level HDPFL.

\begin{table}[t]
    \caption{\revision{Test accuracy ($\%$) of P-FedAvg, DP-FedAvg, IDP-FedAvg, and GDPFed$^+$ on CIFAR-10 with different data heterogeneity degree $\lambda$.}}
    \centering
    \scalebox{0.81}{
    \begin{tabular}{l|cccc|c}
    \toprule
    \textbf{Method} & $\lambda=0.3$ & $\lambda=0.5$ & $\lambda=0.7$ & $\lambda=0.9$ & \textbf{Avg.} \\
    \midrule
    P-FedAvg &
    $\textcolor{black}{49.17}_{\pm 0.58}$ &
    $\textcolor{black}{51.14}_{\pm 0.45}$ &
    $\textcolor{black}{52.55}_{\pm 0.29}$ &
    $\textcolor{black}{53.35}_{\pm 0.55}$ &
    $51.55$ \\
    \midrule
    \midrule
    DP-FedAvg &
    $\textcolor{black}{26.18}_{\pm 1.18}$ &
    $\textcolor{black}{28.66}_{\pm 1.80}$ &
    $\textcolor{black}{29.43}_{\pm 0.92}$ &
    $\textcolor{black}{31.21}_{\pm 0.23}$ &
    $28.87$ \\
    IDP-FedAvg &
    $\textcolor{black}{27.23}_{\pm 1.78}$ &
    $\textcolor{black}{29.93}_{\pm 1.27}$ &
    $\textcolor{black}{31.22}_{\pm 0.46}$ &
    $\textcolor{black}{33.21}_{\pm 0.43}$ &
    $30.40$ \\
    \rowcolor[rgb]{ .9, .9, .9}\textbf{GDPFed$^+$} &
    $\mathbf{\textcolor{black}{28.59}_{\pm 0.71}}$ &
    $\mathbf{\textcolor{black}{32.02}_{\pm 0.57}}$ &
    $\mathbf{\textcolor{black}{34.43}_{\pm 0.94}}$ &
    $\mathbf{\textcolor{black}{35.58}_{\pm 0.78}}$ &
    $\mathbf{32.66}$ \\
    \bottomrule
    \end{tabular}
    }
    \label{tab:heterogeneity_accuracy_avg}
\end{table}

\begin{table}[t]
    \caption{\revision{Test accuracy ($\%$) under different clipping thresholds.}}
    \centering
    \scalebox{0.81}{
    \begin{tabular}{l|l|ccc|c}
        \toprule
        \textbf{Dataset} & \textbf{Method} & $1.0\times$ & $1.25\times$ & $1.5\times$ & \textbf{Average} \\
        \midrule
        \multirow{2}{*}{\revision{FMNIST}} 
            & DP-FedAvg 
              & $\textcolor{black}{71.88}_{\pm 0.15}$ 
              & $\textcolor{black}{68.35}_{\pm 0.60}$ 
              & $\textcolor{black}{63.62}_{\pm 1.23}$ 
              & $\textcolor{black}{67.95}$ \\
            & \cellcolor[rgb]{ .9,  .9,  .9}\textbf{GDPFed$^+$} 
              & \cellcolor[rgb]{ .9,  .9,  .9}$\mathbf{\textcolor{black}{75.83}_{\pm 0.47}}$ 
              & \cellcolor[rgb]{ .9,  .9,  .9}$\mathbf{\textcolor{black}{75.58}_{\pm 0.60}}$ 
              & \cellcolor[rgb]{ .9,  .9,  .9}$\mathbf{\textcolor{black}{74.49}_{\pm 0.81}}$ 
              & \cellcolor[rgb]{ .9,  .9,  .9}$\mathbf{\textcolor{black}{75.30}}$ \\
        \midrule
        \multirow{2}{*}{\revision{SVHN}} 
            & DP-FedAvg 
              & $\textcolor{black}{40.80}_{\pm 1.73}$ 
              & $\textcolor{black}{31.69}_{\pm 2.74}$ 
              & $\textcolor{black}{26.11}_{\pm 2.34}$ 
              & $\textcolor{black}{32.87}$ \\
            & \cellcolor[rgb]{ .9,  .9,  .9}\textbf{GDPFed$^+$} 
              & \cellcolor[rgb]{ .9,  .9,  .9}$\mathbf{\textcolor{black}{71.10}_{\pm 0.58}}$ 
              & \cellcolor[rgb]{ .9,  .9,  .9}$\mathbf{\textcolor{black}{70.63}_{\pm 0.46}}$ 
              & \cellcolor[rgb]{ .9,  .9,  .9}$\mathbf{\textcolor{black}{69.05}_{\pm 0.73}}$ 
              & \cellcolor[rgb]{ .9,  .9,  .9}$\mathbf{\textcolor{black}{70.26}}$ \\
        \midrule
        \multirow{2}{*}{\revision{CIFAR-10}} 
            & DP-FedAvg 
              & $\textcolor{black}{32.10}_{\pm 0.27}$ 
              & $\textcolor{black}{32.11}_{\pm 0.31}$ 
              & $\textcolor{black}{31.10}_{\pm 0.82}$ 
              & $\textcolor{black}{31.77}$ \\
            & \cellcolor[rgb]{ .9,  .9,  .9}\textbf{GDPFed$^+$} 
              & \cellcolor[rgb]{ .9,  .9,  .9}$\mathbf{\textcolor{black}{38.78}_{\pm 0.36}}$ 
              & \cellcolor[rgb]{ .9,  .9,  .9}$\mathbf{\textcolor{black}{38.29}_{\pm 0.69}}$ 
              & \cellcolor[rgb]{ .9,  .9,  .9}$\mathbf{\textcolor{black}{37.50}_{\pm 0.39}}$ 
              & \cellcolor[rgb]{ .9,  .9,  .9}$\mathbf{\textcolor{black}{38.19}}$ \\
        \midrule
        \multirow{2}{*}{\revision{Shakespeare}} 
            & DP-FedAvg 
              & $\textcolor{black}{31.48}_{\pm 2.85}$ 
              & $\textcolor{black}{29.36}_{\pm 2.11}$ 
              & $\textcolor{black}{25.76}_{\pm 0.64}$ 
              & $\textcolor{black}{28.87}$ \\
            & \cellcolor[rgb]{ .9,  .9,  .9}\textbf{GDPFed$^+$} 
              & \cellcolor[rgb]{ .9,  .9,  .9}$\mathbf{\textcolor{black}{42.00}_{\pm 0.45}}$ 
              & \cellcolor[rgb]{ .9,  .9,  .9}$\mathbf{\textcolor{black}{42.61}_{\pm 0.22}}$ 
              & \cellcolor[rgb]{ .9,  .9,  .9}$\mathbf{\textcolor{black}{42.33}_{\pm 0.16}}$ 
              & \cellcolor[rgb]{ .9,  .9,  .9}$\mathbf{\textcolor{black}{42.31}}$ \\
        \bottomrule
    \end{tabular}
    }
    \label{tab:clipping_threshold_accuracy}
\end{table}

\subsubsection{Impact of clipping threshold}  
We now analyze the influence of the clipping threshold \( C \) on model utility. Theoretically, increasing \( C \) results in less aggressive clipping of local model updates, but also amplifies the magnitude of the noise required to satisfy DP constraints.
We evaluate the impact of larger clipping thresholds by scaling the default \( C \) using multiplicative factors: \( 1.25\times \) and \( 1.5\times \). The results, presented in \autoref{tab:clipping_threshold_accuracy}, indicate that as \( C \) increases, the performance of DP-FedAvg degrades significantly, particularly on FMNIST and SVHN datasets. In contrast, our method, GDPFed$^+$, exhibits only minor fluctuations in performance under each setting, demonstrating strong robustness. Notably, on SVHN, GDPFed$^+$ achieves an average test accuracy of \( 70.26\% \), representing a substantial improvement of \( +37.39\% \) over DP-FedAvg. This performance gain is largely attributed to the use of optimized client sampling ratios, which yield more favorable noise multipliers for each privacy group. These results underscore the effectiveness and practical resilience of GDPFed$^+$ under varying clipping thresholds.

\subsubsection{Impact of privacy budget} We investigate how per-group privacy budgets affect the performance of GDPFed$^+$. Specifically, for each dataset, we scale the default budget of each group by multiplicative factors: $0.5\times$, $0.75\times$, $1.0\times$, $1.25\times$, and $1.5\times$.
 For example, under the $0.5\times$ setting, the privacy budgets for FMNIST become $(0.25, 0.75, 1.50)$, which is $ 0.5\times (0.5, 1.5, 3.0)$. The results across all datasets are summarized in \autoref{tab:privacy_budget_accuracy}.

Overall, GDPFed$^+$ consistently outperforms DP-FedAvg across all settings, with particularly notable gains when the privacy budgets are more restrictive (e.g., $0.5\times$ and $0.75\times$). As the scale increases, both methods exhibit improved performance due to the relaxation of privacy constraints, though GDPFed$^+$ maintains a clear advantage throughout. One key observation is that the utility gap between GDPFed$^+$ and DP-FedAvg becomes smaller at larger scales. This is because higher scaling leads to larger per-group privacy budgets, which in turn require less noise to satisfy the privacy guarantees. Consequently, the model utility of GDPFed$^+$ becomes closer to that of DP-FedAvg in such cases.

\begin{table*}[t]
    \caption{Effect of privacy-budget scaling on test accuracy ($\%$).}
    \centering
    \scalebox{1}{
    \begin{tabular}{l|l|ccccc|c}
        \toprule
        \textbf{Dataset} & \textbf{Method} & $0.5\times$ & $0.75\times$ & $1.0\times$ & $1.25\times$ & $1.5\times$ & \textbf{Average} \\
        \midrule
        \multirow{2}{*}{FMNIST} 
            & DP-FedAvg & $\textcolor{black}{57.72}_{\pm 1.83}$ & $\textcolor{black}{68.87}_{\pm 0.55}$ & $\textcolor{black}{71.88}_{\pm 0.15}$ & $\textcolor{black}{73.76}_{\pm 0.39}$ & $\textcolor{black}{74.99}_{\pm 0.29}$ & $\textcolor{black}{69.44}$ \\
            & \cellcolor[rgb]{ .9,  .9,  .9}\textbf{GDPFed$^+$} 
              & \cellcolor[rgb]{ .9,  .9,  .9}$\mathbf{\textcolor{black}{74.87}_{\pm 0.63}}$ 
              & \cellcolor[rgb]{ .9,  .9,  .9}$\mathbf{\textcolor{black}{75.55}_{\pm 0.41}}$ 
              & \cellcolor[rgb]{ .9,  .9,  .9}$\mathbf{\textcolor{black}{75.83}_{\pm 0.47}}$ 
              & \cellcolor[rgb]{ .9,  .9,  .9}$\mathbf{\textcolor{black}{75.97}_{\pm 0.42}}$ 
              & \cellcolor[rgb]{ .9,  .9,  .9}$\mathbf{\textcolor{black}{76.01}_{\pm 0.30}}$ 
              & \cellcolor[rgb]{ .9,  .9,  .9}$\mathbf{\textcolor{black}{75.65}}$ \\
        \midrule
        \multirow{2}{*}{SVHN} 
            & DP-FedAvg & $\textcolor{black}{18.14}_{\pm 0.31}$ & $\textcolor{black}{25.20}_{\pm 1.95}$ & $\textcolor{black}{40.80}_{\pm 1.73}$ & $\textcolor{black}{55.26}_{\pm 1.27}$ & $\textcolor{black}{63.52}_{\pm 1.19}$ & $\textcolor{black}{40.58}$ \\
            & \cellcolor[rgb]{ .9,  .9,  .9}\textbf{GDPFed$^+$} 
              & \cellcolor[rgb]{ .9,  .9,  .9}$\mathbf{\textcolor{black}{57.93}_{\pm 1.55}}$ 
              & \cellcolor[rgb]{ .9,  .9,  .9}$\mathbf{\textcolor{black}{67.54}_{\pm 0.94}}$ 
              & \cellcolor[rgb]{ .9,  .9,  .9}$\mathbf{\textcolor{black}{71.10}_{\pm 0.58}}$ 
              & \cellcolor[rgb]{ .9,  .9,  .9}$\mathbf{\textcolor{black}{72.90}_{\pm 0.73}}$ 
              & \cellcolor[rgb]{ .9,  .9,  .9}$\mathbf{\textcolor{black}{74.03}_{\pm 0.75}}$ 
              & \cellcolor[rgb]{ .9,  .9,  .9}$\mathbf{\textcolor{black}{68.70}}$ \\
        \midrule
        \multirow{2}{*}{CIFAR-10} 
            & DP-FedAvg & $\textcolor{black}{24.20}_{\pm 0.67}$ & $\textcolor{black}{29.63}_{\pm 0.65}$ & $\textcolor{black}{32.10}_{\pm 0.27}$ & $\textcolor{black}{35.06}_{\pm 0.57}$ & $\textcolor{black}{36.89}_{\pm 0.70}$ & $\textcolor{black}{31.58}$ \\
            & \cellcolor[rgb]{ .9,  .9,  .9}\textbf{GDPFed$^+$} 
              & \cellcolor[rgb]{ .9,  .9,  .9}$\mathbf{\textcolor{black}{32.96}_{\pm 0.67}}$ 
              & \cellcolor[rgb]{ .9,  .9,  .9}$\mathbf{\textcolor{black}{36.84}_{\pm 0.12}}$ 
              & \cellcolor[rgb]{ .9,  .9,  .9}$\mathbf{\textcolor{black}{38.78}_{\pm 0.36}}$ 
              & \cellcolor[rgb]{ .9,  .9,  .9}$\mathbf{\textcolor{black}{39.60}_{\pm 0.16}}$ 
              & \cellcolor[rgb]{ .9,  .9,  .9}$\mathbf{\textcolor{black}{40.16}_{\pm 0.11}}$ 
              & \cellcolor[rgb]{ .9,  .9,  .9}$\mathbf{\textcolor{black}{37.27}}$ \\
        \midrule
        \multirow{2}{*}{Shakespeare} 
            & DP-FedAvg & $\textcolor{black}{13.12}_{\pm 8.03}$ & $\textcolor{black}{27.10}_{\pm 1.51}$ & $\textcolor{black}{31.48}_{\pm 2.85}$ & $\textcolor{black}{34.73}_{\pm 0.68}$ & $\textcolor{black}{35.98}_{\pm 0.26}$ & $\textcolor{black}{28.88}$ \\
            & \cellcolor[rgb]{ .9,  .9,  .9}\textbf{GDPFed$^+$} 
              & \cellcolor[rgb]{ .9,  .9,  .9}$\mathbf{\textcolor{black}{37.60}_{\pm 0.57}}$ 
              & \cellcolor[rgb]{ .9,  .9,  .9}$\mathbf{\textcolor{black}{41.11}_{\pm 0.17}}$ 
              & \cellcolor[rgb]{ .9,  .9,  .9}$\mathbf{\textcolor{black}{42.00}_{\pm 0.45}}$ 
              & \cellcolor[rgb]{ .9,  .9,  .9}$\mathbf{\textcolor{black}{42.27}_{\pm 0.59}}$ 
              & \cellcolor[rgb]{ .9,  .9,  .9}$\mathbf{\textcolor{black}{42.55}_{\pm 0.63}}$ 
              & \cellcolor[rgb]{ .9,  .9,  .9}$\mathbf{\textcolor{black}{41.51}}$ \\
        \bottomrule
    \end{tabular}
    }
    \label{tab:privacy_budget_accuracy}
\end{table*}

\begin{table*}[t]
    \caption{Test accuracy ($\%$) of GDPFed$^+$ across datasets under varying sparsification levels; GDPFed-$\mathrm{opc}$ with configuration $(1.0,1.0,1.0)$ serves as the baseline.}
    \centering
    \scalebox{1}{
\begin{tabular}{c|cccc|c}
\toprule
\textbf{Sparsification Level} & \textbf{FMNIST} & \textbf{SVHN} & \textbf{CIFAR-10} & \textbf{Shakespeare} & \textbf{Average} \\
\midrule
$(1.0, 1.0, 1.0)$ &
$\textcolor{black}{75.65}_{\pm 0.53}$ &
$\textcolor{black}{70.85}_{\pm 0.57}$ &
$\textcolor{black}{38.04}_{\pm 0.46}$ &
$\textcolor{black}{41.68}_{\pm 0.44}$ &
$56.56$ \\
\midrule
\midrule
$(0.9, 0.9, 0.9)$ &
$\textcolor{black}{75.89}_{\pm 0.53} \, (\textcolor{black}{+0.24})$ &
$\textcolor{black}{71.04}_{\pm 0.60} \, (\textcolor{black}{+0.19})$ &
$\textcolor{black}{38.69}_{\pm 0.33} \, (\textcolor{black}{+0.65})$ &
$\textcolor{black}{41.94}_{\pm 0.42} \, (\textcolor{black}{+0.26})$ &
$56.89 \, (\textcolor{black}{+0.33})$ \\
$(0.7, 0.7, 0.7)$ &
$\textcolor{black}{75.79}_{\pm 0.44} \, (\textcolor{black}{+0.14})$ &
$\textcolor{black}{70.79}_{\pm 0.41} \, (\textcolor{black}{-0.06})$ &
$\textcolor{black}{38.40}_{\pm 0.42} \, (\textcolor{black}{+0.36})$ &
$\textcolor{black}{41.90}_{\pm 0.34} \, (\textcolor{black}{+0.22})$ &
$56.72 \, (\textcolor{black}{+0.16})$  \\
$(0.5, 0.5, 0.5)$ &
$\textcolor{black}{75.37}_{\pm 0.41} \, (\textcolor{black}{-0.28})$ &
$\textcolor{black}{70.39}_{\pm 0.65} \, (\textcolor{black}{-0.46})$ &
$\textcolor{black}{38.53}_{\pm 0.74} \, (\textcolor{black}{+0.49})$ &
$\textcolor{black}{41.40}_{\pm 0.26} \, (\textcolor{black}{-0.28})$ &
$56.42 \, (\textcolor{black}{-0.14})$  \\
$(0.3, 0.3, 0.3)$ &
$\textcolor{black}{74.92}_{\pm 0.02} \, (\textcolor{black}{-0.73})$ &
$\textcolor{black}{68.94}_{\pm 1.29} \, (\textcolor{black}{-1.91})$ &
$\textcolor{black}{37.50}_{\pm 0.05} \, (\textcolor{black}{-0.54})$ &
$\textcolor{black}{40.33}_{\pm 0.73} \, (\textcolor{black}{-1.35})$ &
$55.42 \, (\textcolor{black}{-1.14})$  \\
$(0.1, 0.1, 0.1)$ &
$\textcolor{black}{72.37}_{\pm 0.35} \, (\textcolor{black}{-3.28})$ &
$\textcolor{black}{61.65}_{\pm 2.10} \, (\textcolor{black}{-9.20})$ &
$\textcolor{black}{35.19}_{\pm 1.58} \, (\textcolor{black}{-2.85})$ &
$\textcolor{black}{36.13}_{\pm 0.12} \, (\textcolor{black}{-5.55})$ &
$51.34 \, (\textcolor{black}{-5.22})$  \\
\midrule
 \rowcolor[rgb]{ .9, .9, .9}$\mathbf{(0.7, 0.8, 0.9)}$ &
$\mathbf{\textcolor{black}{75.83}_{\pm 0.47} \, (\textcolor{black}{+0.18})}$ & $
\mathbf{\textcolor{black}{71.10}_{\pm 0.58} \, (\textcolor{black}{+0.25})} $ &
$\mathbf{\textcolor{black}{38.78}_{\pm 0.36} \, (\textcolor{black}{+0.74})}$ &
$\mathbf{\textcolor{black}{42.00}_{\pm 0.45} \, (\textcolor{black}{+0.32})}$ &
$\mathbf{56.93 \, (\textcolor{black}{+0.37})}$  \\
$(0.5, 0.7, 0.9)$ &
$\textcolor{black}{75.79}_{\pm 0.49} \, (\textcolor{black}{+0.14})$ &
$\textcolor{black}{71.17}_{\pm 0.58} \, (\textcolor{black}{+0.32})$ &
$\textcolor{black}{38.63}_{\pm 0.28} \, (\textcolor{black}{+0.59})$ &
$\textcolor{black}{41.98}_{\pm 0.42} \, (\textcolor{black}{+0.30})$ &
$56.89 \, (\textcolor{black}{+0.33})$  \\
$(0.3, 0.5, 0.7)$ &
$\textcolor{black}{75.62}_{\pm 0.43} \, (\textcolor{black}{-0.03})$ &
$\textcolor{black}{70.83}_{\pm 0.45} \, (\textcolor{black}{-0.02})$ &
$\textcolor{black}{38.69}_{\pm 0.18} \, (\textcolor{black}{+0.65})$ &
$\textcolor{black}{41.83}_{\pm 0.30} \, (\textcolor{black}{+0.15})$ &
$56.74 \, (\textcolor{black}{+0.18})$  \\
$(0.1, 0.3, 0.5)$ &
$\textcolor{black}{75.52}_{\pm 0.27} \, (\textcolor{black}{-0.13})$ &
$\textcolor{black}{70.14}_{\pm 0.76} \, (\textcolor{black}{-0.71})$ &
$\textcolor{black}{38.38}_{\pm 0.77} \, (\textcolor{black}{+0.34})$ &
$\textcolor{black}{40.98}_{\pm 0.38} \, (\textcolor{black}{-0.70})$ &
$56.26 \, (\textcolor{black}{-0.30})$  \\
\bottomrule
\end{tabular}
}
    \label{tab:sparsification_diff}
\end{table*}

\subsubsection{Optimal sparsification levels} Finally, we conduct extensive experiments to investigate how different sparsification levels affect model utility. Specifically, we report the test accuracies of GDPFed-$\mathrm{opc}$ (with the full sparsification level $(1.0, 1.0, 1.0)$ or one can say no sparsification is applied) and GDPFed$^+$ under various sparsification configurations in \autoref{tab:sparsification_diff}. GDPFed-$\mathrm{opc}$ is used as the baseline.

Our results reveal that moderate sparsification levels, such as $(0.9, 0.9, 0.9)$ and $(0.7, 0.8, 0.9)$, can lead to performance improvements across multiple datasets. In contrast, overly aggressive sparsification (e.g., $(0.1, 0.1, 0.1)$) significantly degrades performance, particularly on complex datasets such as SVHN and CIFAR-10. Notably, GDPFed$^+$ with the sparsification level $(0.1, 0.3, 0.5)$ yields only a minor performance drop of $\textcolor{black}{-0.30\%}$ compared to GDPFed-$\mathrm{opc}$, and clearly outperforms configurations like $(0.3, 0.3, 0.3)$ and $(0.1, 0.1, 0.1)$. This supports our intuition that groups with stricter privacy requirements should adopt more aggressive sparsification, while groups with looser privacy constraints can retain more parameters.

Based on these results, we observe that the optimal performance is achieved when GDPFed$^+$ uses the sparsification level $(0.7, 0.8, 0.9)$; therefore, we adopt it as the default configuration in our subsequent experiments. For other datasets not evaluated in this work, we recommend starting with $(0.7, 0.8, 0.9)$ and gradually adjusting the sparsification levels to balance utility and privacy based on task-specific characteristics.

\subsection{\revision{Computational and Communication Overhead}
}
\label{sec: computation_cost}
\revision{
Here, we study the computational and communication overhead introduced by GDPFed$^+$. Specifically, GDPFed$^+$ involves additional operations for solving~\problemautoref{problem:opt} and performing per-group $\mathrm{Top}_k(\cdot)$ sparsification.
The non-convex~\problemautoref{problem:opt} only needs to be solved once before training starts. Empirically, solving this optimization problem takes $0.054\,\mathrm{s}$ on our hardware, which is negligible compared with the overall training process. The per-group $\mathrm{Top}_k(\cdot)$ sparsification involves sorting all parameters within each group’s parameter space, which introduces an additional computational overhead of at most $O(M d \log d)$. Empirically, the sparsification operation takes approximately $0.0007\,\mathrm{s}$ per round in our CIFAR-10 experiments on our hardware, making this overhead negligible in practice. In terms of communication overhead, our method does not introduce additional cost or savings, since the aggregated model parameters remain dense after group-wise aggregation. In conclusion, GDPFed$^+$ introduces only negligible computational overhead on the server and no additional communication overhead, which makes it suitable for IoT settings.
}

\section{Conclusion} \label{sec: conclusion}

In this work, we explore the challenges of achieving client-level DPFL with heterogeneous privacy requirements. Unlike classic methods that must satisfy the strictest privacy requirements across all clients, we propose GDPFed, which partitions clients into groups to ensure group-level DP guarantees. This design preserves
high model utility while accommodating heterogeneous privacy preferences across clients. Based on the privacy and convergence analysis of GDPFed, we introduce GDPFed$^+$, which integrates model sparsification and optimal client sampling ratios to further enhance the utility of GDPFed. GDPFed$^+$ preserves the same privacy guarantees as GDPFed while achieving significant utility improvements, as demonstrated both theoretically and empirically. We discuss promising future directions and the broader impact of our work in Appendix~E.

{
\small
\bibliographystyle{IEEEtran}
\bibliography{ref.bib}
}

\end{document}